\documentclass[fleqn,12pt,twoside]{article}
\usepackage{espcrc1}

\usepackage{graphicx}
\usepackage[figuresright]{rotating}

\newcommand{\AmS}{{\protect\the\textfont2
  A\kern-.1667em\lower.5ex\hbox{M}\kern-.125emS}}

\title{Hypernuclear structure from $\gamma$-ray spectroscopy}

\author{D.J. Millener\address{Brookhaven National Laboratory, 
        Upton, NY 11973, USA}%
        \thanks{Work supported by the US Department
of Energy under contract no. DE-AC02-98CH10886.}}

\begin{document}

\maketitle

\begin{abstract}
 The energies of p-shell hypernuclear $\gamma$ rays obtained from 
recent experiments using the Hyperball at BNL and KEK are used to
constrain the YN interaction which enters into shell-model 
calculations that include  both $\Lambda$ and $\Sigma$ configurations.
\end{abstract}

\section{INTRODUCTION}
 
 At the time of HYP2000 in Torino, it was possible to discuss the
results from two hypernuclear $\gamma$-ray experiments using the 
Hyperball detector array. The first was KEK E419 in which four $\gamma$-ray
transitions in $^7_\Lambda$Li were observed leading to an accurate
identification of three out of four bound excited states of
$^7_\Lambda$Li, in particular the 692-keV separation of the ground-state 
doublet \cite{tamura00}. The lifetime of the 2.05-MeV $5/2^+$
state of $^7_\Lambda$Li was also measured and interpreted in terms
of a shrinkage in the size of the $^6$Li core in $^7_\Lambda$Li
with respect to the free $^6$Li \cite{tanida01}. The second
experiment, BNL E930, measured a small energy difference of
$\sim 31$ keV between the $\sim 3$ MeV $\gamma$-ray transitions from 
the two hypernuclear states based on the lowest $2^+$ state of the 
$^8$Be core in $^9_\Lambda$Be \cite{akikawa02}.

In terms of the standard phenomenological parametrization of the 
$\Lambda$N effective interaction, a good fit to $^7_\Lambda$Li 
can be obtained with  $\Delta =0.48$, S$_\Lambda =-0.01$, S$_{\rm N}
 =-0.40$, and  ${\rm T}=0.03$.
Here, the spin-dependence for a $\Lambda$ in a $0s$ orbit 
interacting with a p-shell core is specified by four radial integrals
$\Delta$, S$_\Lambda$, S$_N$, and T associated with the operators 
$s_N.s_\Lambda$, $l_{N\Lambda}.s_\Lambda$,  
$l_{N\Lambda}.s_N$, and $3(\sigma_N.\widehat{r})
(\sigma_\Lambda.\widehat{r}) -\sigma_N.\sigma_\Lambda$.
For a YNG-type effective interaction, Table 2 of Ref.~\cite{millener01}
shows that the radial integrals take somewhat larger values for the heavier
p-shell hypernuclei because both the nucleon and $\Lambda$ orbits 
become more deeply bound with more confined wave functions despite 
an $A^{1/3}$ increase in the radius of the Woods-Saxon potential wells. 
The substantial value for $\Delta$ is dictated by the ground-state
doublet separation of $^7_\Lambda$Li. The value of
S$_{\rm N}$ is required to bring the excitation energy of the 2.05-MeV
$5/2^+$ state down to its observed value (Table 1 of Ref.~\cite{millener01})
and is also important for the energy of the $1/2^+;1$ state. As far
as S$_{\rm N}$ is concerned there is consistency for the
excitation energies of the $3/2^+$ state in $^{13}_{\ \Lambda}$C
(Table 4 of Ref.~\cite{millener01}) and the excited $1^-$ states
in $^{12}_{\ \Lambda}$C (Table 5 of Ref.~\cite{millener01}). The
values of S$_\Lambda$ and T are constrained to be small by the 
energy separation of the excited-state doublet in $^9_\Lambda$Be
(Table 3 of Ref.~\cite{millener01}) and also play a modest role in the
excitation energy of the $5/2^+$ state and the $7/2^+$, $5/2^+$
separation in $^7_\Lambda$Li.

 In 2001, further $(K^-,\pi^-\gamma)$ experiments were performed
as part of BNL E930 using $^{16}$O and $^{10}$B targets. The primary
purpose of the $^{16}$O experiment was to search for $\sim 6.6$-MeV
$\gamma$ rays from the $\nu p_{3/2}^{-1}\,\Lambda s_{1/2}$ $1^-$
state to the $0^-$ and $1^-$ members of the 
$\nu p_{1/2}^{-1}\,\Lambda s_{1/2}$ ground-state doublet in order
to determine a value for the matrix element of the $\Lambda$N tensor  
interaction which makes a large contribution to the doublet
splitting. This experiment was successful, determining that the $0^-$
state is the ground state and that the doublet separation is 27 keV 
(see the contributions of Ukai and Tamura). In addition, a strong
2.27-MeV $\gamma$ ray and several other candidates were seen in
$^{15}_{\ \Lambda}$N following proton emission from excited states
of $^{16}_{\ \Lambda}$O. The $^{10}$B experiment still shows no
evidence for the ground-state doublet transition in $^{10}_{\ \Lambda}$B.
However, $\gamma$ rays from $^{7}_{\Lambda}$Li and $^9_\Lambda$Be
that provide new information on these hypernuclei were seen.

 Finally, two experiments were performed at KEK in 2002. The first,
KEK E509, searched for and saw hypernuclear $\gamma$ rays following 
stopped $K^-$ interactions on $^7$Li, $^9$Be, $^{10}$B, $^{11}$B, 
and $^{12}$C targets (see the contributions of Miwa and Tamura).
The second, KEK E518, established six $\gamma$-ray transitions in
$^{11}_{\ \Lambda}$B using the 
$^{11}$B($\pi^+,K^+\gamma)^{11}_{\ \Lambda}$B reaction (see the 
contributions of Miura and Tamura).
    
 In the following sections, we investigate how the new information from
experiments with the Hyperball can be understood in terms of
shell-model calculations for p-shell hypernuclei. The one new
ingredient since HYP2000 is that both $\Lambda$ and $\Sigma$
hypernuclear configurations are included so that the explicit
effects of $\Lambda -\Sigma$ coupling are evident, at least for
one model of the YN interaction.

\section{THE EFFECTS OF \boldmath $\Lambda -\Sigma$ COUPLING}
\label{sec:lamsig}

 The coupling of the $\Lambda$N and $\Sigma$N channels is important
and it has long been known that $\Lambda -\Sigma$ coupling makes
an important contribution to the spacing of the $1^+$ and $0^+$ states
of $^4_\Lambda$H and $^4_\Lambda$He. Akaishi et al.
\cite{akaishi00} have given a clear demonstration of this effect
using G-matrices calculated for use in the small model space
of $s$ orbits only. The splittings for the NSC97e and NSC97f 
interactions bracket the observed spacings of the $1^+$ and $0^+$ 
states and it is found that the $\Lambda$N spin-spin interaction and
the $\Lambda -\Sigma$ coupling make comparable contributions to
the spacing. 

 To extend this calculation to p-shell hypernuclei, we take the
YNG interaction of Akaishi for the $\Lambda -\Sigma$ coupling
and perform shell-model calculations with a basis of $p^ns_\Lambda$
and $p^ns_\Sigma$ configurations. Initially, the $\langle N\Lambda
|G|N\Sigma\rangle$ two-body matrix elements have been computed
from the YNG interaction using harmonic oscillator wave functions
with $b = 1.7$ fm. The YNG interaction has non-central components
but the dominant feature is a strong central interaction in the
$^3S$ channel reflecting the second-order effect of the strong
tensor interaction in the $\Lambda$N$ -\Sigma$N coupling. Because
the relative wave function for a nucleon in a $p$ orbit and a
hyperon in an $s$ orbit is roughly half $s$ state and half $p$ state,
the matrix elements coupling $\Lambda$-hypernuclear and   
$\Sigma$-hypernuclear configurations are roughly a factor of two 
smaller than those for the $A=4$ system. Because the energy shifts
for the $\Lambda$-hypernuclear states are given by $v^2/\Delta E$,
where $v$ is the coupling matrix element and $\Delta E\sim 80$ MeV, 
the shifts in p-shell
hypernuclei will be roughly a quarter of those for $A=4$ in
favorable cases; e.g. 150 keV if the  $\Lambda -\Sigma$ coupling
accounts for about half of the $A=4$ splitting. For $T=0$
hypernuclei, the effect will be smaller because the requirement
of a $T=1$ nuclear core for the $\Sigma$-hypernuclear configurations
brings in some recoupling factors which are less than unity.

 We note that many interesting new results have been obtained on
the effects of $\Lambda -\Sigma$ coupling in few-body calculations
for the s-shell hypernuclei \cite{hiyama01,nogga02,nemura02}.

\section{RESULTS FOR P-SHELL HYPERNUCLEI}
\label{sec:spectra}

 We first consider $^7_\Lambda$Li and $^9_\Lambda$Be to illustrate
some features of the inclusion of $\Lambda -\Sigma$ coupling. In
addition, there is new experimental information on each of these
hypernuclei from the BNL E930 run with a $^{10}$B target. Then
we turn to $^{16}_{\ \Lambda}$O and $^{15}_{\ \Lambda}$N which
were studied in the BNL E930 run with a $^{16}$O target and finally
to $^{11}_{\ \Lambda}$B which was studied in the KEK E518 experiment
with a $^{11}$B target.

\subsection{$^{\bf 7}_{\bf\Lambda}$Li}
\label{sec:7li}

  The $\Lambda -\Sigma$ coupling matrix elements for a nucleon in
the p-shell and a hyperon in the s-shell were calculated for the
SC97f(S) effective interaction of Akaishi et al.~\cite{akaishi00}
as described in the preceding section. These matrix elements were
multiplied by 0.9 to simulate the $\Lambda -\Sigma$ coupling of
SC97e(S) and thus the observed doublet splitting for $^4_\Lambda$He
(see \cite{akaishi00}). In the same parametrization as the $\Lambda$N
interaction,  
\begin{equation}
\overline{V}=1.45\quad \Delta =3.04\quad {\rm S}_\Lambda =-0.085\quad 
{\rm S}_{\rm N}  =-0.085\quad {\rm T}=0.157\ .
\label{eq:lamsig}
\end{equation}
 To reproduce the ground-state doublet splitting of 692 keV for
$^7_\Lambda$Li with $\Lambda -\Sigma$ coupling included requires 
a 10\% reduction in $\Delta$ for the $\Lambda$N interaction. 
A small reduction in the magnitude 
of  S$_N$ is then required to fit the excitation energy of the 
$5/2^+$ state exactly. The contributions to the energy spacings for 
\begin{equation}
 \Delta =0.432\quad {\rm S}_\Lambda =-0.010\quad {\rm S}_{\rm N}
 =-0.390\quad {\rm T}=0.028\ 
\label{eq:7li}
\end{equation}
are given in Table~\ref{tab:7li}. The coefficents of the $\Lambda$N 
interaction parameters are very close to those given in 
Ref.~\cite{millener01}. The actual energy shifts due to
$\Lambda -\Sigma$ coupling for the $1/2^+$, $3/2^+$, $5/2^+$ and
$7/2^+$ states with $T=0$ are 77, 6, 74, and 0 keV while those
for the $1/2^+$, $3/2^+$, and $5/2^+$ states with $T=1$ are 97,
101, and 95 keV.

 Taking simple LS wave functions for the $^6$Li core, one can predict
quite well the effects of $\Lambda -\Sigma$ coupling in the full
shell-model calculation. For the ground-state, the only
important $\Sigma$ configuration involves a $0^+$, $T=1$ core. The
recoupling brings in an extra factor of $1\sqrt{3}$ in the coefficient
before the $^3S$ G-matrix element compared with the $A=4$ $0^+$ state.
The net effect is an order of magnitude smaller shift than for the
\begin{table}[ht]
\caption{Contributions from $\Lambda -\Sigma$ coupling and the
spin-dependent components of the effective $\Lambda$N interaction
to energy spacings in $^7_\Lambda$Li. Energies are in keV.}
\begin{tabular*}{\textwidth}{@{}c@{\extracolsep{\fill}}cccccccc}
\hline
Level pair & $\Delta E_{core}$ &  $\Lambda\Sigma$ & $\Delta$ & 
 S$_\Lambda$ & S$_{\rm  N}$ &  T & $\Delta E$ & Expt.  \\
\hline
 $3/2^+ - 1/2^+$ & ~~~0 & 71 &  621  & $-1$  & $-6$ & $-6$ & 690 & 692  \\
 $5/2^+ - 1/2^+$ & 2186 & 3 &  75  & 11  & $-272$ & 31  &  2052 & 2050 \\
 $1/2^+ - 1/2^+$ & 3565 & $-20$ &  415  & 0  & $-183$ & $-2$  &  3773 & 3877\\
 $7/2^+ - 5/2^+$ & ~~~0 & 74 &  559  & $-22$  & $-7$ & $-67$  &  511 & 470\\
\hline
\end{tabular*}
\label{tab:7li}
\end{table}
$A=4$ system. It is easy to see that the $\Lambda -\Sigma$ coupling
for the $3/2^+$ member of the ground-state doublet occurs only
through the much weaker non-central interactions. Thus, 
$\Lambda -\Sigma$ coupling provides only 10\% of the ground-state
doublet splitting compared with $\sim 50$\% for $A=4$. Comprehensive
few-body calculations for the $A=4$ and $A=7$ hypernuclei should be
able to put a strong constraint on the overall strength of the
$\Lambda -\Sigma$ coupling in the free YN interaction.

 The energy shifts for the doublet based on the $3^+$ state of the core
are similar to those in the ground-state doublet while the energy
shifts for the $T=1$ hypernuclear states are somewhat larger than
for the $1/2^+$ and $5/2^+$ $T=0$ states because the $\Lambda$ and
$\Sigma$ configurations can have the same $T=1$ cores and the 
$\Sigma$ configurations can also have $T=0$ cores.

 One new piece of information on the $^7_\Lambda$Li spectrum has been 
obtained from BNL E930. The 2050-keV $5/2^+\to 1/2^+$ $\gamma$ ray
is seen rather strongly from the unbound region of $^{10}_{\ \Lambda}$B
produced in the $^{10}$B($K^-,\pi^-$) reaction. A $\gamma$-ray line
at 470 keV is seen in coincidence with the 2050-keV $\gamma$ ray and
has been interpreted as the $7/2^+\to 5/2^+$ transition (Tamura in
these proceedings). The calculated energy in Table~\ref{tab:7li}
of 511 keV is somewhat too high and can be accommodated by
small increases in the magnitudes of S$_\Lambda$ and/or T. Such
increases are constrained by the excited-state doublet splitting
in $^9_\Lambda$Be (see Table~\ref{tab:9be}). 

We note that the 692-keV 
$3/2^+\to 1/2^+$ $\gamma$ ray is also seen in BNL E930 and that
the 2050-keV $5/2^+\to 1/2^+$ $\gamma$ ray is seen strongly in
KEK E509 via the (stopped $K^-,\gamma$) reaction on $^{10}$B.

\subsection{$^{\bf 9}_{\bf\Lambda}$Be}
\label{sec:9be}

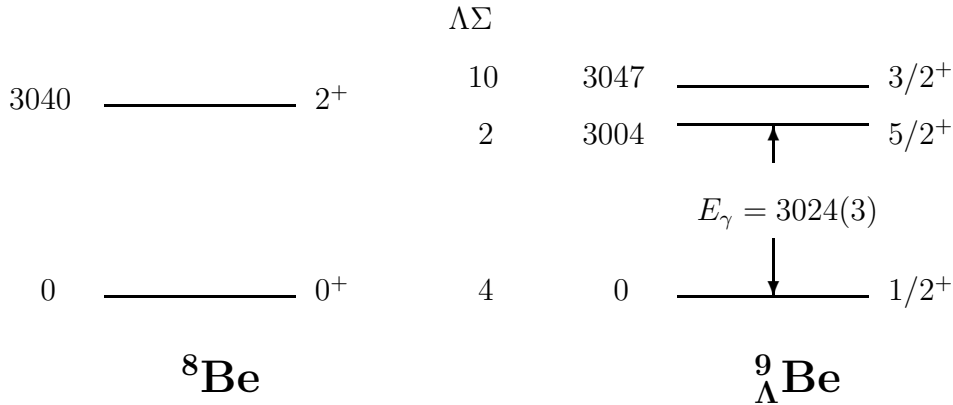
\begin{figure}[bh]
\begin{center}
\setlength{\unitlength}{1in}
\thicklines
\begin{picture}(5,2.5)
\put(0.5,1.0){\line(1,0){1.0}}
\put(0.5,2.0){\line(1,0){1.0}}
\put(3.5,1.0){\line(1,0){1.0}}
\put(3.5,1.9){\line(1,0){1.0}}
\put(3.5,2.1){\line(1,0){1.0}}
\put(0.0,0.98){~~~0}
\put(0.0,1.98){3040}
\put(1.6,0.98){$0^+$}
\put(1.6,1.98){$2^+$}
\put(3.0,0.98){~~~0}
\put(3.0,1.8){3004}
\put(3.0,2.1){3047}
\put(4.6,0.98){$1/2^+$}
\put(4.6,1.8){$5/2^+$}
\put(4.6,2.1){$3/2^+$}
\put(2.3,2.4){$\Lambda\Sigma$}
\put(2.4,0.98){~4}
\put(2.4,1.8){~2}
\put(2.4,2.1){10}
\put(3.6,1.4){$E_\gamma = 3024(3)$}
\put(4.0,1.3){\vector(0,-1){0.3}}
\put(4.0,1.7){\vector(0,1){0.2}}
\put(0.9,0.5){\Large $^{\bf 8}${\bf Be}}
\put(3.9,0.5){\Large $^{\bf 9}_{\bf \Lambda}${\bf Be}}
\end{picture}
\end{center}
\vspace*{-1.4cm}
\caption{Energy levels of $^9_\Lambda$Be and the $^8$Be core. The 
small shifts due to $\Lambda -\Sigma$ coupling are shown in the 
center. The measured $\gamma$-ray energies are 3024(3) and
3067(3) keV with a separation of 43(5) keV.}
\label{fig:be9}
\end{figure}

 The bound-state spectrum for $^9_\Lambda$Be is shown in Fig.~\ref{fig:be9},
which also gives the $\gamma$-ray energies from a reanalysis of the
BNL E930 data \cite{akikawa02}, for the parameter set
\begin{equation}
 \Delta =0.557\quad {\rm S}_\Lambda =-0.013\quad {\rm S}_{\rm N}
 =-0.549\quad {\rm T}=0.038\ . 
\label{eq:be9}
\end{equation}

\begin{table}[ht]
\caption{Contributions from $\Lambda -\Sigma$ coupling and the
spin-dependent components of the effective $\Lambda$N interaction
to the doublet spacing in $^{9}_{\Lambda}$Be. Energies are in keV.
The spectrum is shown on the right hand side of Fig.~\ref{fig:be9}.
The second line gives the coefficients of each 
$\Lambda$N interaction parameter in MeV.}
\begin{tabular*}{\textwidth}{@{}c@{\extracolsep{\fill}}rrrrrrr}
\hline
Level pair & $\Delta E_{core}$ &  $\Lambda\Sigma$ & $\Delta$ & 
 S$_\Lambda$ & S$_{\rm  N}$ &  T & $\Delta E$   \\
\hline
  $3/2^+ - 5/2^+$ & 0 & $-8$ & $-20$ & 32 & $-1$ & 38 & 43  \\
 & & & $-0.037$ & $-2.464$ & 0.003 & 0.994 & \\
\hline
\end{tabular*}
\label{tab:9be}
\end{table}
The breakdown of the doublet splitting is given in Table~\ref{tab:9be}.
As can be seen, the contribtions of S$_\Lambda$ and T work against
those from $\Delta$ and the $\Lambda -\Sigma$ coupling. Certainly,
the value of S$_\Lambda$ cannot be large. The parameter set chosen
puts the $3/2^+$ state above the $5/2^+$ state but the order is not
determined by this experiment. However, in the 2001 run of BNL E930
on a $^{10}$B target, only the upper level is seen following
$^{10}_{\ \Lambda}$B$\to {^9_\Lambda}$Be$+p$. Then, we can deduce
\Large
\begin{figure}[bh]
\vspace*{1.2cm}
\begin{center}
\setlength{\unitlength}{0.7in}
\thicklines
\begin{picture}(6,5)(0.0,0.0)
\put(1.0,1.0){\line(1,0){1.}}
\put(1.0,2.5){\line(1,0){1.}}
\put(1.0,3.05){\line(1,0){1.}}
\put(2.1,1.0){2.00}
\put(2.1,2.5){5.00}
\put(2.1,3.05){6.40}
\put(0.0,1.1){$^9_\Lambda{\rm Be}+p$}
\put(0.0,3.05){$^6_\Lambda{\rm Li}+\alpha$}
\put(3.0,5.5){\vector(-2,-3){1.66}}
\put(1.7,4.0){$\alpha$}
\put(0.0,2.2){$^9_\Lambda{\rm Be}(\frac{3}{2}^+,\frac{5}{2}^+)+p$}
\put(3.0,5.5){\vector(-1,-2){1.5}}
\put(3.0,3.25){\vector(-2,-1){1.5}}
\put(2.5,2.8){$p$}
\put(2.5,4.2){$p$}
\put(3.5,0.0){\line(1,0){1.}}
\put(3.5,0.2){\line(1,0){1.}}
\put(3.5,1.2){\line(1,0){1.}}
\put(3.5,1.3){\line(1,0){1.}}
\put(3.5,3.2){\line(1,0){1.}}
\put(3.5,3.35){\line(1,0){1.}}
\put(3.5,5.4){\line(1,0){1.}}
\put(3.5,5.55){\line(1,0){1.}}
\put(3.0,-0.02){~~~0}
\put(3.0,0.2){0.22}
\put(3.0,1.1){2.47}
\put(3.0,1.3){2.70}
\put(3.0,3.1){6.51}
\put(3.0,3.35){6.70}
\put(3.0,5.3){10.82}
\put(3.0,5.55){11.10}
\put(4.6,-0.02){$1^-$}
\put(4.6,0.2){$2^-$}
\put(4.6,1.1){$2^-$}
\put(4.6,1.3){$3^-$}
\put(4.6,3.1){$3^-$}
\put(4.6,3.35){$4^-$}
\put(4.6,5.3){$3^-$}
\put(4.6,5.55){$4^-$}
\put(5.3,0.2){[0.584]}
\put(5.3,1.1){[0.109]}
\put(5.3,1.3){[0.495]}
\put(5.3,3.1){[0.222]}
\put(5.3,3.35){[0.122]}
\put(5.3,5.3){[1.042]}
\put(5.3,5.55){[0.063]}
\put(3.8,-0.5){\LARGE $^{\bf 10}_{\bf\ \Lambda}${\bf B}}
\put(6.0,0.2){[0.576]}
\put(6.0,1.1){[0.123]}
\put(6.0,1.3){[0.478]}
\put(6.0,3.1){[0.567]}
\put(6.0,3.35){[0.115]}
\put(6.0,5.3){[0.697]}
\put(6.0,5.55){[0.043]}
\end{picture}
\end{center}
\caption{Proton decay of $^{10}_{\ \Lambda}$B to $^9_\Lambda$Be.
Formation strengths for non-spin flip production in the $(K^-,\pi^-)$
reaction are given on the right for two p-shell models. Thresholds
for particle decay of the $^{10}_{\ \Lambda}$B states are given on the
left.}
\label{fig:p9be}
\end{figure}
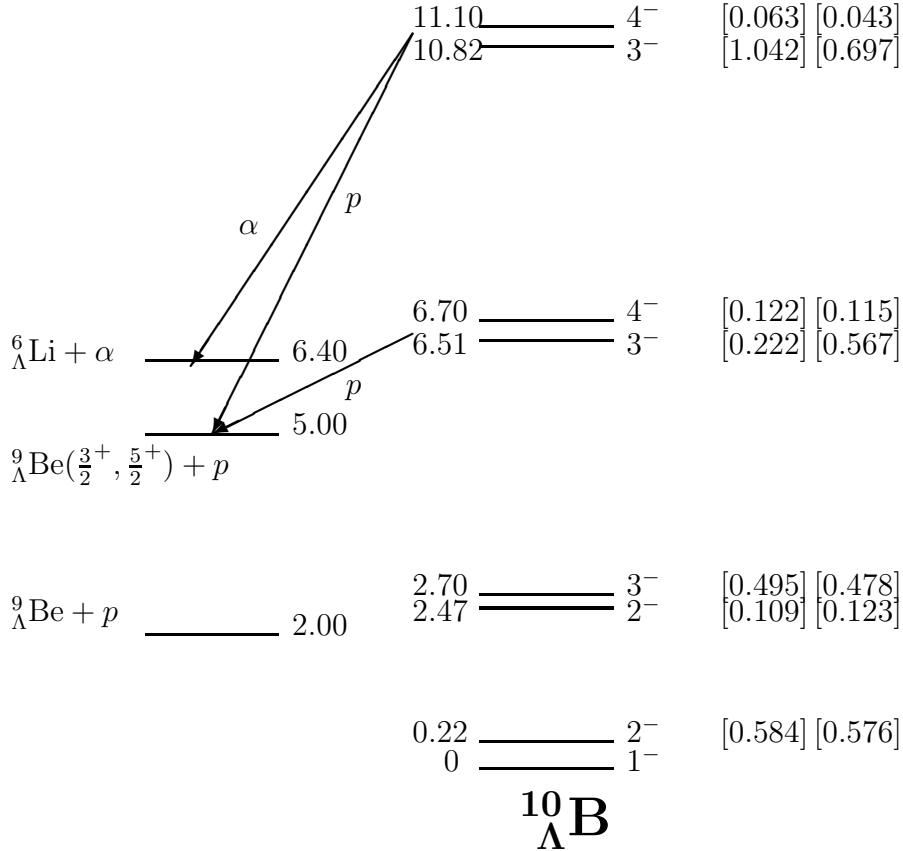
\normalsize
that the $3/2^+$ state is the upper member of the doublet via the 
following reasoning.
Four states of $^9$B are reached strongly by neutron removal
from $^{10}$B \cite{kull68} and the hypernuclear doublets based on
these states are shown in Fig.~\ref{fig:p9be}. The structure
factors which govern the population of these states are given at
the right of the figure for two p-shell interactions. The relative
neutron pickup strength to the two $7/2^-$ states which give rise to the 
$3^-$/$4^-$ doublets above the $^9_\Lambda{\rm Be}^* +p$ threshold
is very sensitive to the non-central components of the p-shell
interaction. Formation of the the $3^-$ states is favored for the 
dominant $p_{3/2}$ removal by the coupling to get $\Delta L =1$
and $\Delta S =0$. The proton decay arises from $^9{\rm B}(7/2^-)\to
{^8{\rm Be}(2^+) + p}$ in the core. The $4^-$ states proton decay
to $^9_\Lambda{\rm Be}(5/2^+)$ and from the recoupling
$(2^+\times p_{3/2})7/2^-\times s_\Lambda\to  (2^+\times s_\Lambda)J_f
\times p_{3/2}$ one finds that the $3^-$ states proton decay to 
the $3/2^+$ and $5/2^+$ states in the ratio of 32 to 3. Overall, the
the $3/2^+$ state is favored by a factor of more than 3. The only
caveat to this argument is that the uppermost $3^-$ state doesn't 
$\alpha$ decay too much.

\subsection{$^{\bf 10}_{\ \bf\Lambda}$B}
\label{sec:10b}

 Table~\ref{tab:10b} shows details of the energy splitting of the
ground-state doublet of $^{10}_{\ \Lambda}$B for the
$^7_\Lambda$Li parameter set given in Eq.~(\ref{eq:7li}). The
predicted doublet splitting should be observable but was not
seen in an early Brookhaven experiment \cite{chrien90} and is not
seen in BNL E930. Theoretically, the splitting is mainly due to
the $\Lambda$N spin-spin interaction but does have the interesting
feature that the effect of $\Lambda -\Sigma$ coupling works
against the $\Lambda$N spin-spin interaction. This happens because
the coupling matrix element to the $^9{\rm B}(gs)\times s_\Sigma$
configuration is much larger for the $2^-$ state than for the
$1^-$ state; in fact, the coupling to the $^9{\rm B}(1/2^-)\times 
s_\Sigma$ state is most important for the $1^-$ state.

\begin{table}[ht]
\caption{Contributions from $\Lambda -\Sigma$ coupling and the
spin-dependent components of the effective $\Lambda$N interaction
to energy spacings in $^{10}_{\ \Lambda}$B. Coefficients (first line)
in MeV, energies  in keV. The energy shifts due to $\Lambda -\Sigma$ 
coupling are 49 and 34 keV for the $2^-$ and $1^-$ states.}
\begin{tabular*}{\textwidth}{@{}c@{\extracolsep{\fill}}ccccc}
\hline
 $\Lambda\Sigma$ & $\Delta$ & S$_\Lambda$ & S$_{\rm  N}$ &  T & 
$\Delta E$   \\
\hline
   & $0.579$ & $1.413$ & 0.013 & $-1.073$ & \\
 $-15$ &  $250$  &  $-14$  &  $-5$  &  $-32$  &  180 keV   \\
\hline
\end{tabular*}
\label{tab:10b}
\end{table}

\subsection{$^{\bf 16}_{\ \bf\Lambda}$O}
\label{sec:16o}

\begin{figure}[th]
\vspace*{0.6cm}
\setlength{\unitlength}{1.0in}
\thicklines
\begin{picture}(6,2.45)(-0.3,0.0)
\put(0.5,0.0){\line(1,0){1.5}}
\put(0.5,1.4){\line(1,0){1.5}}
\put(0.5,1.5){\line(1,0){1.5}}
\put(0.5,2.0){\line(1,0){1.5}}
\put(0.5,2.4){\line(1,0){1.5}}
\put(0.5,2.5){\line(1,0){1.5}}
\put(0.0,-0.05){~~~0}
\put(0.0,1.32){5183}
\put(0.0,1.52){5241}
\put(0.0,2.0){6176}
\put(0.0,2.35){6793}
\put(0.0,2.52){6857}
\put(2.1,-0.05){$1/2^-$}
\put(2.1,1.32){$1/2^+$}
\put(2.1,1.52){$5/2^+$}
\put(2.1,2.00){$3/2^-$}
\put(2.1,2.35){$3/2^+$}
\put(2.1,2.52){$5/2^+$}
\put(2.9,1.5){$\Lambda\Sigma$}
\put(2.9,-0.05){~27}
\put(2.9,0.2){~57}
\put(2.9,2.30){~87}
\put(2.9,2.49){~~6}
\put(4.0,0.0){\line(1,0){1.5}}
\put(4.0,0.3){\line(1,0){1.5}}
\put(4.0,2.29){\line(1,0){1.5}}
\put(4.0,2.44){\line(1,0){1.5}}
\put(3.5,-0.02){~~~0}
\put(3.5,0.2){~~28}
\put(3.5,2.24){6554}
\put(3.5,2.44){6873}
\put(5.6,-0.02){$0^-$}
\put(5.6,0.2){$1^-$}
\put(5.6,2.24){$1^-$}
\put(5.6,2.44){$2^-$}
\put(4.4,2.27){\circle*{0.1}}
\put(4.3,2.4){71}
\put(5.1,2.27){\circle*{0.1}}
\put(5.0,2.4){29}
\put(4.2,1.0){6554}
\put(4.4,0.9){\vector(0,-1){0.9}}
\put(4.4,1.2){\line(0,1){1.09}}
\put(4.9,1.0){6526}
\put(5.1,0.9){\vector(0,-1){0.6}}
\put(5.1,1.2){\line(0,1){1.09}}
\put(4.4,-0.5){\LARGE $^{\bf {16}}_{\ \bf \Lambda}${\bf O}}
\put(1.0,-0.5){\LARGE $^{\bf {15}}${\bf O}}
\end{picture}
\vspace*{0.5cm}
\caption{Spectra of $^{15}$O and the $p^{-1}s_\Lambda$ states of
$^{16}_{\ \Lambda}$O for the parameter set of Eq.~(\ref{eq:16o}).
The energy shifts due to $\Lambda -\Sigma$ coupling are given in keV.} 
\label{fig:16o}
\end{figure}
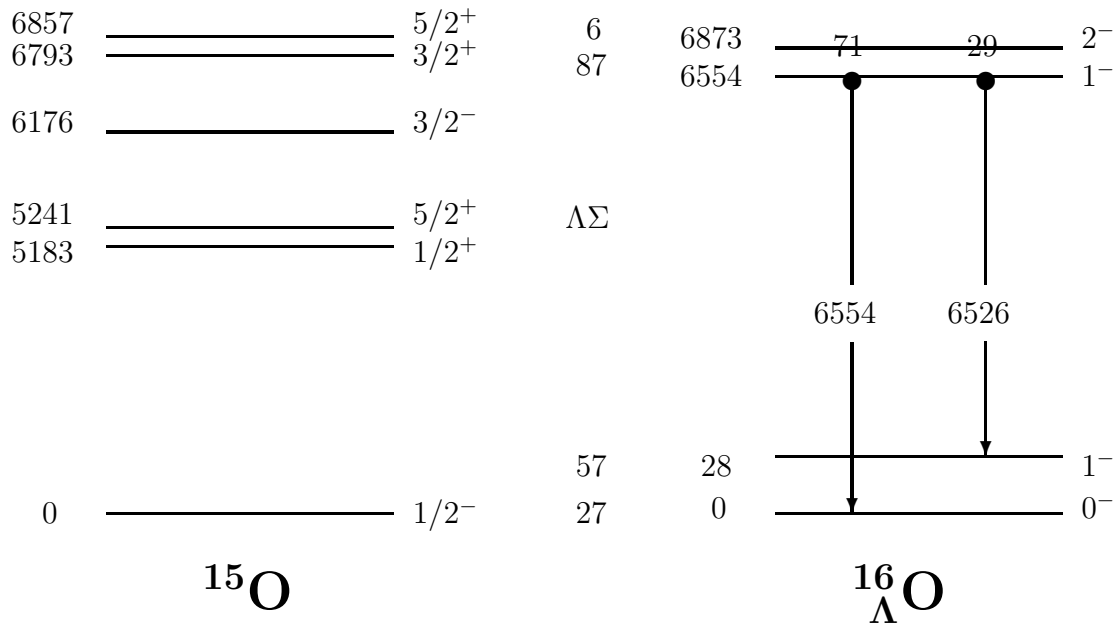

 A measurement of the ground-state doublet splitting in $^{16}_{\ \Lambda}$O
has long been desirable because of the strong dependence of the
energy separation on the $\Lambda$N tensor interaction. This objective
has been achieved in BNL E930 (Ukai and Tamura, these proceedings) by
observing the $\gamma$ rays from the excited $1^-$ level 
(see Fig.~\ref{fig:16o}). The measured $\gamma$-ray energies and intensities
are $6558.6\pm 1.4$ keV ($200\pm 23$ counts) and $6532.1\pm 1.8$ keV
($128\pm 20$ counts) for a splitting of $26.5\pm 2.3$ keV. Fig.~\ref{fig:16o}
shows the calulated spectrum for the parameter set
\begin{equation}
 \Delta =0.468\quad {\rm S}_\Lambda =-0.011\quad {\rm S}_{\rm N}
 =-0.354\quad {\rm T}=0.030\ . 
\label{eq:16o}
\end{equation}

 Table~\ref{tab:16o10} shows that the dominant contributions to
the ground-state doublet splitting are due to $\Delta$ and T.
The rise of the $\gamma$-ray energies from the 6176 keV separation
of the core levels in $^{15}$O is of course due to S$_N$ which
contributes 530 keV to the separation of the two $1^-$ levels.
The fact that the energy separation between the two lowest
peaks in a $(\pi^+,K^+)$ experiment is $6.22\pm 0.06$ MeV 
\cite{hashimoto01} suggests that hypernuclear levels based on the lowest
positive-parity levels contribute to the cross section.
\begin{table}[hb]
\caption{Breakdown of the ground-state doublet splitting for
$^{16}_{\ \Lambda}$O.} 
\begin{tabular*}{\textwidth}{@{}c@{\extracolsep{\fill}}ccccc}
\hline
 $\Lambda\Sigma$ & $\Delta$ & S$_\Lambda$ & S$_{\rm  N}$ &  T & 
$\Delta E$  \\
\hline
   & $-0.382$ & $1.378$ & $-0.004$ & 7.850 & \\
 $-30$ &  $-179$  &  $-15$  &  $1$  &  $235$  &  28 keV   \\
\hline
\end{tabular*}
\label{tab:16o10}
\end{table}

\subsection{$^{\bf 15}_{\ \bf\Lambda}$N}
\label{sec:15n}

 In the $^{16}{\rm O}(K^-,\pi^-)^{16}_{\ \Lambda}{\rm O}$ reaction
used for BNL E930, $p^{-1}p_\Lambda$ $0^+$ states are strongly 
excited at about 10.6 and 17.0 MeV in excitation energy along
with a broad distribution of $s^{-1}s_\Lambda$ strength centered
near 25 MeV \cite{bruckner78}. These levels can decay by proton
emission (the threshold is at $\sim 7.8$ MeV) to $^{15}_{\ \Lambda}$N
via $s^4p^{10}(sd)s_\Lambda$ components in their wave functions. For
example, the wave function of the pure non-spurious component
underlying the s-hole strength is
\begin{equation}
|s^{-1}s_{\Lambda};0^+\rangle = \sqrt{4/5} s^3p^{12}s_\Lambda  
 +\sqrt{1/5} s^4p^{10}(sd)s_\Lambda\ .
\label{eq:shole}
\end{equation}
\begin{figure}[th]
\setlength{\unitlength}{0.9in}
\thicklines
\begin{picture}(4,5)(-1.0,0.2)
\put(0.5,0.0){\line(1,0){1.5}}
\put(0.5,0.3){\line(1,0){1.5}}
\put(0.5,2.44){\line(1,0){1.5}}
\put(0.5,4.20){\line(1,0){1.5}}
\put(0.5,4.87){\line(1,0){1.5}}
\put(0.0,-0.02){~~~0}
\put(0.0,0.3){~110}
\put(0.0,2.42){2398}
\put(0.0,4.18){4071}
\put(0.0,4.85){4874}
\put(2.1,-0.02){$3/2^+$}
\put(2.1,0.3){$1/2^+$}
\put(2.1,2.42){$1/2^+$;1}
\put(2.1,4.18){$1/2^+$}
\put(2.1,4.85){$3/2^+$}
\put(0.7,4.93){99.5}
\put(1.1,2.44){\circle*{0.1}}
\put(1.05,2.5){92}
\put(1.4,4.20){\circle*{0.1}}
\put(1.3,4.26){100}
\put(1.7,2.44){\circle*{0.1}}
\put(1.65,2.5){8}
\put(3.1,0.3){$\tau\sim 87$ ps}
\put(3.1,2.42){$\tau\sim 0.46$ ps}
\put(3.1,4.18){$\tau\sim 5.2$ fs}
\put(3.1,4.85){$\tau\sim 1.6$ fs}
\put(1.1,1.1){\vector(0,-1){1.1}}
\put(1.1,1.4){\line(0,1){1.04}}
\put(0.9,1.2){2398}
\put(4.9,1.2){$\Lambda\Sigma$}
\put(4.9,-0.02){~56}
\put(4.9,0.3){~12}
\put(4.9,2.42){102}
\put(4.9,4.18){~74}
\put(4.9,4.63){~10}
\put(0.6,3.2){2476}
\put(0.8,3.4){\line(0,1){1.47}}
\put(0.8,3.1){\vector(0,-1){0.66}}
\put(1.2,3.2){1673}
\put(1.4,3.1){\vector(0,-1){0.66}}
\put(1.4,3.4){\line(0,1){0.80}}
\put(1.5,1.2){2288}
\put(1.7,1.1){\vector(0,-1){0.8}}
\put(1.7,1.4){\line(0,1){1.04}}
\end{picture}
\caption{States of $^{15}_{\ \Lambda}$N based on the $1^+;0$ ground state,
2313 keV $0^+;1$ state, and 3948 keV $1^+;0$  state of the $^{14}$N core
for the parameter set of Eq.~(\ref{eq:16o}).}
\label{fig:15n}
\end{figure}
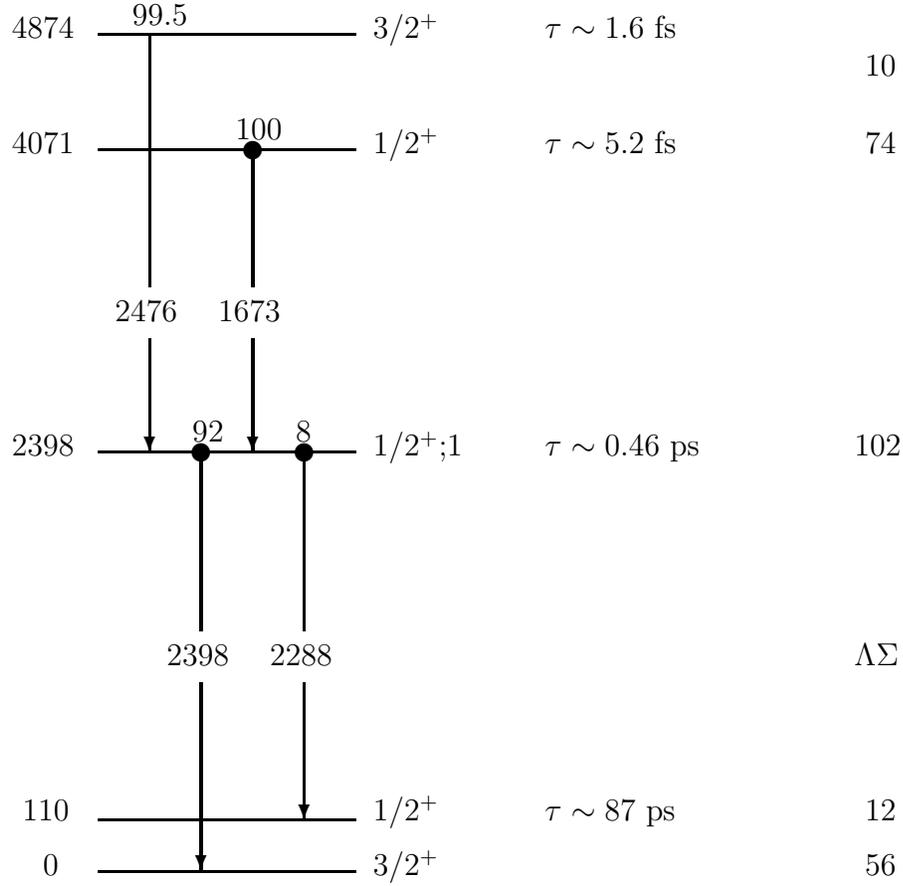
By setting a cut for small pion angles, where the $\Delta L =0$ cross 
section is large, it was possible to observe $\gamma$ rays from states
of $^{15}_{\ \Lambda}$N in BNL E930 (Tamura in these proceedings).
The strongest line is observed at 2268 keV in the spectra without
Doppler correction. In Fig.~\ref{fig:15n}, this line would be
identified with the $1/2^+;1\to 3/2^+$ transition. Note that the
ordering of states in the ground-state doublet violates the usual
ordering that has the lower-spin state lowest; in the simplest
model with $p_{1/2}$ nucleon holes only, the $1/2^+$ state would be 
lowest and the $^{15}_{\ \Lambda}$N splitting would be 1.5 times
that in $^{16}_{\ \Lambda}$O \cite{millener85}.
 As for $^{16}_{\ \Lambda}$O, $\Delta$ and T give large and
cancelling contributions to the ground-state doublet splitting.
However, because of the deviation of the $^{14}$N ground-state
wave function from $jj$ coupling, the coefficient of $\Delta$
in Table~\ref{tab:15n} for $^{15}_{\ \Lambda}$N is relatively 
larger than that of T and this leads to the inversion.

\begin{table}[hb]
\caption{Breakdown of the ground-state doublet splitting in
$^{15}_{\ \Lambda}$N or the parameter set of Eq.~(\ref{eq:16o}).} 
\begin{tabular*}{\textwidth}{@{}c@{\extracolsep{\fill}}ccccc}
\hline
 $\Lambda\Sigma$ & $\Delta$ & S$_\Lambda$ & S$_{\rm  N}$ &  T & 
$\Delta E$  \\
\hline
   & $0.756$ & $-2.250$ & $0.035$ & $-9.864$ & \\
$44$ &  $354$  &  $25$  &  $-12$  &  $-296$  &  110 keV \\
\hline
\end{tabular*}
\label{tab:15n}
\end{table}

 In the weak-coupling limit, the B(M1) values for the decay to the
$3/2^+$ and $1/2^+$ states are in the ratio of 2:1. However, the
core M1 transition is very weak and mostly orbital because the 
$\langle\sigma\tau\rangle$ matrix element is very small in analogy to 
$^{14}$C $\beta$ decay. Then, small admixtures of 
$1^+_2\times s_\Lambda$ configurations into
the final states produce substantial cancellations 
\begin{eqnarray}
 M(1/2^+;1\to 1/2^+) & = & -\left\{(0.9980)(0.9992)(-0.25496) +
  (0.0619)(0.9992)(3.21472)\right\} \nonumber \\
 & = & -\left\{ -0.25427 + 0.19883\right\} = 0.0554 \ ,
\label{eq:mix11}
\end{eqnarray}
\begin{eqnarray}
 M(1/2^+;1\to 3/2^+) & = & \left\{(0.9980)(0.9992)(-0.25496) +
  (0.0302)(0.9992)(3.21472)\right\} \nonumber \\
 & = & \left\{ -0.25427 + 0.09700\right\} = -0.1573 \ .
\label{eq:mix13}
\end{eqnarray}
The first numbers in the second product are the admixing amplitudes
for the $1^+_2\times s_\Lambda$ components while the third numbers
in each product are the core M1 matrix elements. The full result 
for an effective M1 operator which reproduces local p-shell M1 data
gives a partial lifetime of 0.5 ps (for $E_\gamma = 2263$ keV)
and the branching ratios shown in Fig.~\ref{fig:15n}. Clearly, the
observation of both $\gamma$-rays and the comparison with the
ground-state doublet splitting of $^{16}_{\ \Lambda}$O would be
very informative.

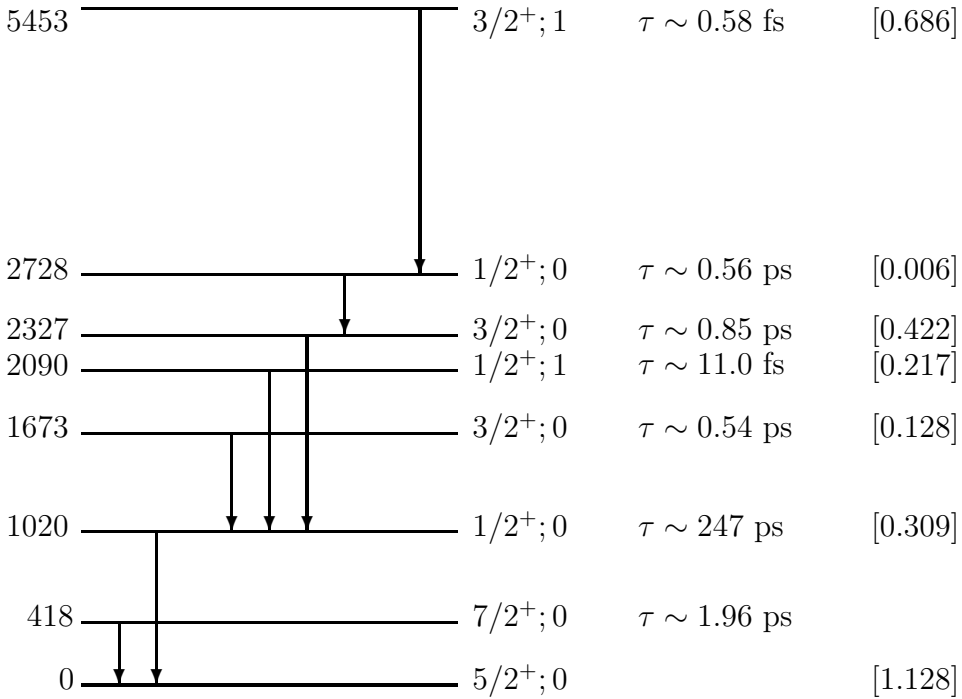
\begin{figure}[bh]
\vspace*{-0.3cm}
\setlength{\unitlength}{1cm}
\thicklines
\begin{picture}(16,12)(0,0)
\put(2.0,2.0){\line(1,0){5.}}
\put(1.7,1.95){0}
\put(7.2,1.95){$5/2^+;0$}
\put(12.5,1.95){[1.128]}
\put(2.0,2.836){\line(1,0){5.}}
\put(1.0,2.786){~~418}
\put(7.2,2.786){$7/2^+;0$}
\put(2.5,2.836){\vector(0,-1){0.836}}
\put(9.4,2.786){$\tau\sim 1.96$ ps}
\put(2.0,4.040){\line(1,0){5.}}
\put(1.0,3.990){1020}
\put(7.2,3.990){$1/2^+;0$}
\put(12.5,3.990){[0.309]}
\put(9.4,3.990){$\tau\sim 247$ ps}
\put(3.0,4.040){\vector(0,-1){2.040}}
\put(2.0,5.346){\line(1,0){5.}}
\put(1.0,5.296){1673}
\put(7.2,5.296){$3/2^+;0$}
\put(12.5,5.296){[0.128]}
\put(4.0,5.346){\vector(0,-1){1.306}}
\put(9.4,5.296){$\tau\sim 0.54$ ps}
\put(2.0,6.180){\line(1,0){5.}}
\put(1.0,6.130){2090}
\put(7.2,6.130){$1/2^+;1$}
\put(12.5,6.130){[0.217]}
\put(4.5,6.180){\vector(0,-1){2.140}}
\put(9.4,6.130){$\tau\sim 11.0$ fs}
\put(2.0,6.654){\line(1,0){5.}}
\put(1.0,6.604){2327}
\put(7.2,6.604){$3/2^+;0$}
\put(12.5,6.604){[0.422]}
\put(5.0,6.654){\vector(0,-1){2.614}}
\put(9.4,6.604){$\tau\sim 0.85$ ps}
\put(2.0,7.456){\line(1,0){5.}}
\put(1.0,7.406){2728}
\put(7.2,7.406){$1/2^+;0$}
\put(12.5,7.406){[0.006]}
\put(5.5,7.456){\vector(0,-1){0.802}}
\put(9.4,7.406){$\tau\sim 0.56$ ps}
\put(2.0,11.0){\line(1,0){5.}}
\put(1.0,10.7){5453}
\put(7.2,10.7){$3/2^+;1$}
\put(12.5,10.7){[0.686]}
\put(9.4,10.7){$\tau\sim 0.58$ fs}
\put(6.5,11.0){\vector(0,-1){3.544}}
\end{picture}
\vspace*{-2.6cm}
\caption{Partial level scheme for $^{11}_{\ \Lambda}$B together
with calculated lifetimes and formation strengths. The indicated
$\gamma$-ray transitions are those that are predicted to be strong
enough to be seen in KEK E419.}
\label{fig:11b}
\end{figure}

\subsection{$^{\bf 11}_{\ \bf\Lambda}$B}
\label{sec:11b}

 The $^{10}$B core has a number of p-shell levels at low energy 
and the proton threshold in $^{11}_{\ \Lambda}$B is at 7.72 MeV
which leads to a very rich spectrum of $\gamma$-ray transitions
that potentially provides for many cross checks on the parametrization
of the YN interaction. Fig.~\ref{fig:11b} shows a partial spectrum 
for $^{11}_{\ \Lambda}$B and gives predictions for $\gamma$-rays 
that might be seen via the $^{11}$B($\pi^+,K^+\gamma)^{11}_{\ \Lambda}$B 
reaction used for KEK E518. The Barker I interaction \cite{barker81} 
used for the $^{10}$B core was chosen to reproduce the empirically
optimized wave functions discussed by Kurath~\cite{kurath79}. With
standard p-shell effective g-factors and charges these wave functions
give a good description of electromagnetic transitions in $^{10}$B.
For the hypernuclear calculation, the formation strengths for each
level below the proton threshold were combined with the calculated
$\gamma$-ray branches to predict the relative intensities of $\gamma$-rays
produced in the cascade. The lowest $1/2^+;0$ level serves as a
collection point although for the predicted energy and lifetime in 
Fig.~\ref{fig:11b} this level would weak decay more than 50\% of the time.
In the data, a 1482 keV $\gamma$ ray is the strongest and five other
$\gamma$-rays at 262, 454, 500, 564, and 2479 keV are observed with
intensities of $9 - 17$\% relative the 1482-keV line. At an energy
of 1482 keV, the lifetime of the $1/2^+_1;0$ level should be $\sim 38$ ps
and the level should mostly $\gamma$ decay. The 2479 keV shows up
in the Doppler corrected spectrum and it is natural to identify it
with the strong M1 transition from the the $3/2^+;1$ level to the 
$1/2^+_2;0$ level. The other four observed transitions are more difficult 
to place especially in light of the discrepancy between theory
and experiment for the $1/2^+_1;0$ level and the fact that the predictions
for the $1/2^+$ and $3/2^+$ levels based on the $1^+;0$ core levels
at 718 and 2154 keV are rather volatile with respect to changes in
the p-shell interaction.

\section{CONCLUSION}

 We are now able to understand considerable body of hypernuclear
data in terms of one parametrization of the YN interaction used in
shell-model calculations which include $\Lambda$ and $\Sigma$ degrees
of freedom although the ground-state doublet of $^{10}_{\ \Lambda}$B 
is still a problem. 


\end{document}